\documentclass[prd,11pt,showpacs,showkeys]{revtex4-1}

\usepackage[utf8]{inputenc}
\usepackage{textcomp}
\usepackage{bm}
\usepackage{bm}
\usepackage{amsmath}    
\usepackage{amssymb} 
\usepackage{graphicx}   
\usepackage{verbatim}   
\usepackage{color}      
\usepackage{subfigure}  
\usepackage{hyperref}   
\definecolor{amaranth}{rgb}{0.9, 0.17, 0.31}
\definecolor{purple(munsell)}{rgb}{0.62, 0.0, 0.77}
\definecolor{americanrose}{rgb}{1.0, 0.01, 0.24}
\definecolor{palatinateblue}{rgb}{0.15, 0.23, 0.89}
\definecolor{royalblue(web)}{rgb}{0.25, 0.41, 0.88}
\definecolor{hanpurple}{rgb}{0.32, 0.09, 0.98}
\definecolor{beaublue}{rgb}{0.74, 0.83, 0.9}
\definecolor{carminered}{rgb}{1.0, 0.0, 0.22}
\definecolor{brightpink}{rgb}{1.0, 0.0, 0.5}
\definecolor{vividviolet}{rgb}{0.62, 0.0, 1.0}

\hypersetup{ linktoc=all,
    colorlinks, linkcolor={palatinateblue},
    citecolor={brightpink}, urlcolor={amaranth}}

\newcommand{\be}{\begin{equation}}
\newcommand{\ee}{\end{equation}}
\newcommand{\bs}{\begin{split}} 
\newcommand{\bea}{\begin{eqnarray}}
\newcommand{\eea}{\end{eqnarray}}

\begin{document}

\title{Thermal properties of Klein-Gordon Oscillator in the Context of Amelino-Camelia and Magueijo-Smolin Doubly Special Relativity (DSR) frameworks } 

\author{Abdelmalek Boumali$^1$ \footnote{\email{abdelmalek.boumali@univ-tebessa;boumali.abdelmalek@gmail.com (Corr. Author)}}, Nosratollah Jafari$^2$ \footnote{\email{  nosrat.jafari@fai.kz }}, Bekdaulet Shukirgaliyev$^3$ \footnote{\email{ Bekdaulet.shukirgaliyev@nu.edu.kz}}, Fadila Serdouk$^1$ \footnote{\email{  fadila.serdouk@univ-tebessa.dz }}
\smallskip 
\\
$^1$ Laboratory of theoretical and applied Physics Echahid Cheikh Larbi Tebessi University, Algeria
\\
$^2$ Fesenkov Astrophysical Institute, 050020, Almaty, Kazakhstan
\\
$^2$ Al-Farabi Kazakh National University, Al-Farabi av. 71, 050040 Almaty, Kazakhstan
\\
 $^2$ Center for Theoretical Physics, Khazar University, 41 Mehseti Street, Baku, AZ1096, Azerbaijan
\\ 
 $^3$ Physics Department, Nazarbayev University, Astana, 010000, Kazakhstan
\\
$^3$ Energetic Cosmos Laboratory, Nazarbayev University, 010000 Astana, Kazakhstan
\\
$^3$ K. Zhubanov Aktobe Regional University, Aktobe, 030000, Kazakhstan
}
     
\begin{abstract}
We examine the thermal and statistical properties of the one-dimensional Klein–Gordon oscillator within two prominent Doubly Special Relativity (DSR) frameworks: Amelino–Camelia and Magueijo–Smolin. Using the modified dispersion relations specific to each formulation, we derive the positive energy spectra, construct the partition function via the Euler–Maclaurin method, and compute key thermodynamic quantities, including the specific heat Cv, as functions of temperature and the deformation scale. Planck-scale corrections produce distinct, theoretically resolvable shifts in both the position and magnitude of the Cv peak in the two models. An accompanying entropy analysis reveals that these peaks correspond to smooth Schottky-type anomalies: the specific heat curves remain analytic and positive across the explored temperature range, and thus do not indicate latent or continuous thermodynamic phase transitions. These comparative results provide a robust diagnostic framework for differentiating DSR prescriptions in relativistic quantum systems and reinforce the transition-free character of their thermal response.

\end{abstract}

\maketitle

\tableofcontents

\section{Introduction}
DSR theories represent an essential extension of Einstein’s special relativity by introducing a second invariant scale in addition to the speed of light $c$, namely the Planck energy $ E_p = \sqrt{\hbar c^5 / G} \approx 10^{19} \, \text{GeV}$. Traditional special relativity preserves only the invariance of the speed of light across inertial frames. In contrast, DSR modifies the relativistic energy-momentum relations to incorporate quantum gravitational effects that become significant near the Planck scale. Two notable models within this framework are the Amelino-Camelia DSR \cite{Amelino-Camelia:2000stu,Amelino-Camelia:2002uql} and the Magueijo-Smolin (MS) DSR \cite{Magueijo:2001cr}, each proposing unique modifications that preserve observer independence of both $c$ and $E_p$ \cite{JafariPhysRevD}.
In addition, while both models aim to incorporate a second invariant scale—the Planck energy $E_{P}$—their formulations and physical implications differ substantially. The Amelino--Camelia (AC) framework arises from deformations of the Lorentz algebra, often realized via $\kappa$-Poincaré symmetries, and is closely linked to non-commutative geometry. This approach modifies the momentum sector, leading to altered dispersion relations and potentially an energy-dependent speed of light, with phenomenological consequences such as time-of-flight delays for high-energy photons from distant astrophysical sources. By contrast, the Magueijo--Smolin (MS) model is constructed through nonlinear reparametrizations of momentum space that preserve both $c$ and $E_{P}$ as invariants. In practice, this deformation primarily affects the mass term at first order in the Planck length, producing distinctive corrections to particle rest energies and threshold conditions in high-energy reactions.
The coexistence of these two variants reflects different theoretical motivations: AC-type modifications are motivated by quantum-gravity-inspired momentum-space curvature, whereas MS-type modifications focus on preserving exact invariance of both $c$ and $E_{P}$ within a relativistic kinematic framework. Physically, they may be viewed as complementary descriptions, each probing different aspects of Planck-scale kinematics. Potential avenues for distinguishing them experimentally include high-precision astrophysical timing measurements, which are more sensitive to AC-type dispersion effects, and collider or cosmic-ray threshold tests, which could reveal MS-type mass-sector deformations. Although current constraints remain many orders of magnitude away from Planck-scale sensitivity, these scenarios provide concrete targets for future observational and experimental programs \cite{Amelino-Camelia:2000stu,Amelino-Camelia:2002uql, Magueijo:2001cr, JafariPhysRevD}.
These theoretical frameworks offer a phenomenological avenue for investigating how Planck‑scale physics can modify fundamental quantum systems \cite{JafariPhysLettB2020,JafariPhysLettB2024,JafariPhysLettB2025,JafariPhysRevD,GuvendiDSRPhysLettB2024,Coraddu:2009sb}. 
Among such systems, relativistic quantum oscillators occupy a privileged position. The relativistic harmonic oscillator is particularly noteworthy because it is one of the few quantum mechanical models that admit an exact analytical solution.
The Dirac oscillator (DO)—the relativistic extension of the harmonic oscillator—constitutes an interaction potential of both theoretical and practical importance. First introduced by Ito et al \cite{ito1967}, the DO arises from replacing the canonical momentum $\vec{p}$ in the Dirac equation with $\vec{p}-i m \beta \omega \vec{r}$, where $\vec{r}$ is the position vector, $m$ the particle mass, and $\omega$ the oscillator frequency. Moshinsky and Szczepaniak subsequently revived interest in the model and coined the term "Dirac oscillator" \cite{moreno1989}, noting that its non‑relativistic limit reduces to a harmonic oscillator augmented by a pronounced spin–orbit coupling.
From a physical standpoint, the DO can be interpreted as describing the interaction between an anomalous magnetic moment and a linearly varying electric field \cite{moreno1989,martinez1992}. The corresponding electromagnetic potential was derived explicitly by Benitez et al. \cite{Benitez1990}. Because it remains exactly solvable within the Dirac formalism—and owing to its broad applicability in fields such as quantum optics, nuclear physics, and condensed‑matter analogues—the Dirac oscillator continues to attract significant research attention \cite{quesne1990,Quimbay2013a,Quimbay2013b,Boumaliejtp2015,Boumalizna12015,Boumalizna22015,boumali2020}.
Most recently, Franco‑Villafane et al. \cite{Franco-Villafane2013} achieved the first experimental realization of the one‑dimensional Dirac oscillator using a microwave setup, thereby providing a tangible platform for probing its characteristic vibrational modes.
The Klein–Gordon oscillator (KGO), describing spin-0 particles, was originally proposed by Bruce and Manning \cite{bruce1993} through the introduction of a linear interaction term into the Klein–Gordon equation, effectively modifying the canonical momentum operator in a manner analogous to minimal coupling. This transformation extends the conventional harmonic oscillator into a fully relativistic framework for bosonic systems, closely paralleling the role of the Dirac oscillator (DO) for spin-1/2 particles \cite{moshinsky1989}. As one of the few exactly solvable models in relativistic quantum mechanics, the KGO provides a powerful theoretical platform for investigating phenomena such as Planck-scale corrections, dynamics in curved spacetimes, and other deformations inspired by quantum gravity.
Bruce and Manning \cite{bruce1993} introduced the Klein-Gordon oscillator (KGO) by using the replacements
\be
\boldsymbol{P}\rightarrow\boldsymbol{P}-im\hat{\gamma}\hat{\Omega}\boldsymbol{Q},\qquad\boldsymbol{P}=-i\frac{\partial}{\partial\boldsymbol{q}},
\ee
where,
\be 
\boldsymbol{P}=\hat{\eta}\boldsymbol{p},\quad\boldsymbol{Q}=\hat{\eta}\boldsymbol{q},
\ee
with $\hat{\eta}^{2}=1$.  Similar approaches can be found in \cite{Dvoeglazo1994, taketani1940, feshbach1958, Boumali2023}.\\
Understanding how DSR-induced modifications affect the energy spectrum and thermodynamic properties of the KGO can reveal important insights into the phenomenology of quantum gravity. It may help identify potential experimental signatures at high energies. The Klein-Gordon oscillator (KGO) is an ideal theoretical laboratory to explore the interplay between quantum mechanics, relativity, and gravity-induced deformations.\\
This work aims to investigate the thermal and statistical properties of the one-dimensional Klein-Gordon oscillator within the context of the Amelino-Camelia and Magueijo-Smolin Doubly Special Relativity theories. By deriving the modified Klein-Gordon equations and solving for their energy spectra, we compute the partition functions and key thermodynamic quantities such as the specific heat. This analysis highlights how the deformation parameter associated with the Planck scale influences the system's behavior, with a particular emphasis on the positive energy sector ensured by the Foldy-Wouthuysen transformation.

Throughout, we evaluate the equilibrium thermodynamic quantities using only the positive energy branch of the spectrum. The Klein–Gordon equation is cast in the two-component Feshbach–Villars form to identify positive/negative sectors; for the free or interaction problem, this enables standard Foldy-Wouthuysen (FW) block diagonalization (see \cite{KhanJagannathan1994OperatorApproach,Silenko2008}). We emphasize that FW/FV here serve solely as kinematic tools for sector identification and decoupling in the single-particle setting. Within this single-particle framework, we compare the thermal behavior of the Klein–Gordon oscillator under the Amelino–Camelia (AC) and Magueijo–Smolin (MS) DSR deformations.
\\
The paper is structured as follows. Section II introduces the one-dimensional Klein–Gordon oscillator and its formulation within the Doubly Special Relativity framework. Section III examines the statistical and thermodynamic properties, including analytical approximations of the partition function and specific heat, for both the Amelino–Camelia and Magueijo–Smolin models. Section IV summarizes the main conclusions and discusses perspectives for future research.

\section{Klein--Gordon Oscillator under DSR Frameworks}
Before addressing the implications of Planck-scale deformations, it is instructive to recall the standard formulation of the one-dimensional Klein-Gordon oscillator (KGO).
\subsection*{A. Standard Klein--Gordon Oscillator}
The free Klein--Gordon equation is given by
\begin{equation} \label{eq:1}
\left(\square - \frac{E^{2} - m^{2} c^{4}}{c^{2}}\right) \phi(x) = 0,
\end{equation}
where $\square$ denotes the d'Alembert operator, explicitly defined as
\begin{equation} \label{eq:2}
\square = \frac{1}{c^{2}} \frac{\partial^{2}}{\partial t^{2}} - \frac{\partial^{2}}{\partial x^{2}} .
\end{equation}
To introduce an interaction analogous to the Dirac oscillator, one typically employs the non-minimal substitution
\begin{equation} \label{eq:3}
p_x \rightarrow p_x - i m \omega x,
\end{equation}
where $\omega$ denotes the oscillator frequency. Applying this substitution, the Klein-Gordon equation transforms into
\begin{equation}\label{eq:4}
\left[c^{2} (p_x + i m \omega x)(p_x - i m \omega x) - E^{2} + m^{2} c^{4}\right] \phi(x) = 0.
\end{equation}
This expression can be rearranged into the form of a harmonic oscillator equation:
\begin{equation} \label{eq:5}
\left(\frac{p_x^{2}}{2m} + \frac{m\omega^{2}}{2} x^{2}\right) \phi(x)
= \left(\frac{m c^{2} \hbar \omega + E^{2} - m^{2} c^{4}}{2 m c^{2}}\right) \phi(x)
\equiv \tilde{E} \, \phi(x),
\end{equation}
where $\tilde{E}$ is defined as
\begin{equation} \label{eq:6}
\tilde{E} = \frac{m c^{2} \hbar \omega + E^{2} - m^{2} c^{4}}{2 m c^{2}}.
\end{equation}
Here, $\Phi(x,t) = e^{-i E t} \phi(x)$. The resulting equation corresponds precisely to the standard one-dimensional harmonic oscillator. Consequently, the eigenvalue spectrum is given by
\begin{equation} \label{eq:7}
\epsilon_{n} = \pm m c^{2} \sqrt{1 + 2 \frac{\omega}{mc^2} n}, 
\end{equation}
The spectrum exhibits symmetric branches for positive and negative energies, consistent with relativistic quantum oscillator dynamics.
\subsection*{B. Amelino--Camelia DSR Framework}
In the Amelino--Camelia (AC) formulation of DSR \cite{JafariPhysLettB2024}, the modified Klein-Gordon equation of order $\mathcal{O}(E^3/E_p^3)$ can be expressed by separating the standard Klein--Gordon (KG) part from the deformation term:
\begin{equation}   
-\frac{1}{c^2}\frac{\partial^2}{\partial t^2}\Phi + \frac{i}{E_p}\frac{\partial}{\partial t}\vec{\nabla}^2\Phi - \left(m^2c^2 - \vec{\nabla}^2\right)\Phi = 0. \label{eq3}  
\end{equation}  
or
\begin{equation}  
\underbrace{-\frac{1}{c^2}\frac{\partial^2}{\partial t^2}\Phi - \left(m^2c^2 - \vec{\nabla}^2\right)\Phi}_{\text{Standard KG form}}
\;+\;
\underbrace{\frac{i}{E_p}\frac{\partial}{\partial t}\,\vec{\nabla}^2\Phi}_{\text{AC deformation}}
= 0.
\label{eq3.1}
\end{equation}
The first term corresponds to the usual relativistic scalar wave equation, while the second term represents the Planck-scale deformation.
Introducing the one-dimensional Klein-Gordon oscillator via the non-minimal substitution
\[
p_x \rightarrow p_x - i m \omega x, \quad (\hbar = 1),
\]
yields
\begin{equation}
E^{2} \Phi =
\left[ m^{2}c^{4}
+ c^2 \left( 1 + \frac{i}{E_p}\frac{\partial}{\partial t} \right)
p^{2} \right] \Phi,
\label{eq5}
\end{equation}
with
\begin{equation}
p^{2} =
\left( p_{x} + i m \omega x \right)
\left( p_{x} - i m \omega x \right).
\label{eq6}
\end{equation}
After rearrangement, the harmonic oscillator form appears explicitly:
\begin{equation}
\left( \frac{p^{2}}{2m} + \frac12 m \omega^{2} x^{2} \right) \Phi
= \left[
\frac{E^{2} - m^{2}c^{4}}
{2mc^2\left(1 + \frac{E}{E_p}\right)}
+ \frac{\omega}{2}
\right] \Phi.
\label{eq9}
\end{equation}
Here, the deformation enters through the $(1 + E/E_p)^{-1}$ factor.
Solving this harmonic oscillator problem gives the following.
\begin{equation}
E_n =
\underbrace{\frac{m c^2\omega}{E_p} n}_{\text{linear AC shift}}
\;\pm\;
m c^2 \sqrt{
\left( \frac{\omega}{E_p} \right)^{2} n^{2}
+ 1 + 2 \frac{\omega}{m c^{2}} n
}.
\label{eq12}
\end{equation}
The first term is a pure AC deformation proportional to $1/E_p$, absent in the standard KG oscillator.
In the undeformed limit $E_p \to \infty$, the deformation terms vanish and the expression reduces to
\begin{equation}
E_n \to \pm m c^{2} \sqrt{ 1 + 2 \frac{\omega}{m c^{2}} n },
\label{eq13}
\end{equation}
Restoring the familiar relativistic Klein--Gordon oscillator spectrum \cite{Boumaliejtp2015}.
\subsection*{C. Magueijo--Smolin DSR Framework}
In the Magueijo--Smolin (MS) DSR model \cite{Coraddu:2009sb}, the modified one-dimensional Klein--Gordon equation can be expressed as
\begin{equation}
\underbrace{\left( -\frac{1}{c^2}\frac{\partial^2}{\partial t^2} + \vec{\nabla}^2 \right)\Phi}_{\text{Standard KG form}}
=
m^2 c^2
\underbrace{\left(1 - \frac{i}{E_p} \frac{\partial }{\partial t} \right)^2}_{\text{MS deformation}}
\Phi.
\label{eq14}
\end{equation}
The left-hand side contains the standard KG operator, while the right-hand side introduces the Planck-scale modification via $(1 - i\,\partial_t/E_p)^2$.
Applying the non-minimal substitution
\[
p_x \rightarrow p_x - i m \omega x,
\]
and assuming $\hbar = 1$, we obtain the following after rearranging:
\begin{equation}
\left( \frac{p^{2}}{2m} + \frac12 m \omega^{2} x^{2} \right) \Phi
=
\left[
-\frac{m c^2}{2}
\left(1 - \frac{E}{E_p} \right)^{2}
+ \frac{\omega}{2} + \frac{E^{2}}{2 m c^{2}}
\right] \Phi.
\label{eq15}
\end{equation}
The harmonic oscillator quantization condition is then
\begin{equation}
-\frac{m c^2}{2} \left(1 - \frac{E}{E_p} \right)^{2} + \frac{\omega}{2} + \frac{E^{2}}{2 m c^{2}}
= \omega \left( n + \frac12 \right).
\label{eq16}
\end{equation}
Solving for $E_n$ gives
\begin{equation}
E_n =
\frac{E_p\, m^{2} c^{4}}
{m^{2} c^{4} - E_p^{2}}
\;\pm\;
\frac{E_p\, m c^{2} \sqrt{E_p^{2} + \frac{2 n \omega}{m c^{2}}\left(E_p^{2} - m^{2} c^{4}\right)}}
{m^{2} c^{4} - E_p^{2}}.
\label{eq17}
\end{equation}
Here, the deformation parameter $E_p$ appears in both the numerator and the denominator, making the modification more intricate than in the AC case.
In the undeformed limit $E_p \to \infty$, all deformation-dependent fractions vanish, and the spectrum reduces to
\begin{equation}
E_n \to \pm m c^{2} \sqrt{ 1 + 2 \frac{\omega}{m c^{2}} n },
\end{equation}
recovering the standard relativistic Klein--Gordon oscillator spectrum \cite{Boumaliejtp2015}.
\subsection*{D. Comparative Discussion}
The two DSR-inspired formulations produce qualitatively distinct modifications to the Klein-Gordon oscillator spectrum, reflecting the fundamentally different ways in which the Planck-scale deformation is implemented.
In the \emph{Amelino-Camelia} (AC) framework, the deformation enters through an energy-dependent rescaling of the kinetic term. This results in an energy spectrum of the form
\begin{equation}
E_{n}^{\text{AC}} = \frac{m c^{2} \omega}{E_{P}} n \pm m c^{2} \sqrt{\left( \frac{\omega}{E_{P}} \right)^{2} n^{2} + 1 + 2 \frac{\omega}{m c^{2}} n},
\end{equation}
which features both a linear shift in $n$ and a modified square root dependence. The linear term, proportional to $n/E_{P}$, is absent in the standard relativistic case and dominates only when the oscillator quantum number or the frequency is sufficiently large relative to the Planck scale. As $E_{P} \to \infty$, the expression reduces smoothly to the conventional Klein--Gordon oscillator spectrum.
In contrast, the \emph{Magueijo--Smolin} (MS) formulation implements the deformation through a nonlinear modification of the mass term, yielding the following results.
\begin{equation}
E_{n}^{\text{MS}} = \frac{E_{P} m}{2 c^{4} \left(m^{2} c^{4} - E_{P}^{2}\right)}
\left( m^{2} c^{4} - E_{P}^{2} \pm \sqrt{E_{P}^{2} + \frac{2 n \omega}{m c^{2}} \left(E_{P}^{2} - m^{2} c^{4}\right)} \right).
\end{equation}
Here, $E_{P}$ appears both in the numerator and in the denominator, producing a more intricate dependence on the deformation scale. Unlike in the AC case, there is no explicit additive term linear in $n$; instead, the deformation modifies the curvature of the spectrum in a nonlinear fashion. 
Both models exhibit symmetric positive- and negative-energy branches, preserving the particle--antiparticle structure characteristic of relativistic quantum oscillators. In the limit $E_{P} \to \infty$, both reduce to
\begin{equation}
E_{n} \to \pm m c^{2} \sqrt{1 + 2 \frac{\omega}{m c^{2}} n},
\end{equation}
recovering the standard Klein-Gordon oscillator result.\\
From a phenomenological standpoint, the differences between the two models are significant: the AC deformation predicts a linear-in-$n$ correction to the spectrum at high energies, while the MS deformation alters the dispersion in a non-polynomial way. These distinctions, though suppressed far below the Planck scale, could in principle manifest in systems where relativistic oscillator modes are probed at extreme energies, such as in certain condensed matter analogues or early-universe field configurations.\\
The following section explores the thermodynamic implications of these spectra by evaluating the corresponding partition functions and specific heats. This analysis provides a macroscopic perspective on how Planck-scale modifications could be indirectly probed.

\section{Statistical Properties}

Before proceeding with the calculation of the thermal properties of the modified Klein-Gordon oscillator, we summarize the key findings from the previous section:

\begin{itemize}
    \item  Both Eqs. (\ref{eq12}) and (\ref{eq17}) form the basis for evaluating thermodynamic quantities such as the partition function and specific heat of the oscillator.
    \item The distinctions between these equations can, in principle, provide observable signatures to differentiate various quantum gravity-inspired modifications of special relativity, as explored further in the analysis and figures of this work.
\end{itemize}
In essence, Eq. (\ref{eq12}) and Eq. (\ref{eq17}) represent the fundamental results for the energy spectra of the one-dimensional Klein-Gordon oscillator within two prominent DSR frameworks. They illustrate how Planck-scale effects alter quantum systems, with each DSR model imparting a unique modification to the energy levels and, consequently, to the thermodynamic behavior of the system.

In this section, our main goal is to determine the statistical properties of the system. To this end, we focus on calculating the partition function $Z$. As described in \cite{boumali2015a}, the partition function is given by\cite{boumali2015a,Boumaliejtp2015,Boumalizna12015,Boumalizna22015}:
\begin{equation}
Z = 1 + \sum_{n=1}^{\infty} e^{-\beta E_{n}}.
\label{eq:zp}
\end{equation}
Although the partition function is naturally expressed as an infinite sum, directly evaluating this sum can be computationally intensive or even intractable, especially for complex systems. To overcome these challenges, we employ the powerful Euler-Maclaurin formula, a well-established technique in mathematical analysis and extensively discussed in the literature (see \cite{Andrews1999}). The Euler-Maclaurin formula bridges the gap between discrete sums and continuous integrals, providing a systematic way to approximate sums by integrals and correction terms involving derivatives and Bernoulli numbers or polynomials.
The general form of the Euler-Maclaurin formula is as follows (see \cite{Andrews1999}):
\begin{equation}
\sum_{n=a}^{b} f(n) = \frac{1}{2} \left(f(b) + f(a)\right) + \int_{a}^{b} f(n)\, dn + \sum_{i=2}^{p} \frac{B_i}{i!} \left( f^{(i-1)}(b) - f^{(i-1)}(a) \right) - \int_{a}^{b} \frac{B_p(1-t)}{p!} f^{(p)}(t)\, dt,
\label{eq:fz}
\end{equation}
where $a$ and $b$ are integer numbers such that $b-a$ is a positive integer, $B_n$ are Bernoulli numbers, $B_p$ are Bernoulli polynomials and $p$ is a positive integer. The function $f$ must have a continuous derivative $p$. The notation $\{x\}$ denotes the fractional part of $x$. The remainder (or error) term is given by:
\begin{equation}
R_k = \int_{a}^{b} \frac{B_p(1-t)}{p!} f^{(p)}(t)\, dt.
\label{eq:rz}
\end{equation}
The importance of the Euler-Maclaurin formula lies in its ability to provide highly accurate approximations to sums, especially when the function $f(x)$ and its derivatives decay rapidly as $x$ increases. When $f(x)$ and its derivatives vanish at infinity, the formula simplifies as shown in \cite{SiouaneLow2024, SiouaneTheo2024}:
\begin{equation}
\sum_{n=0}^{\infty} f(n) = \frac{f(0)}{2} + \int_{0}^{\infty} f(n)\, dn + \sum_{i=2}^{p} \frac{B_i\, f^{(i-1)}(0)}{i!} - \int_{0}^{\infty} \frac{B_p(1-t)}{p!} f^{(p)}(t)\, dt,
\label{eq:sz}
\end{equation}
The initial Bernoulli numbers are $B_0=1$, $B_1=-\frac{1}{2}$, $B_2=\frac{1}{6}$, and $B_4=-\frac{1}{30}$, with all odd Bernoulli numbers beyond $B_1$ being zero. The Bernoulli polynomials $B_n(x)$ are defined by the generating function \cite{Andrews1999}:
\begin{equation}
\frac{t e^{t x}}{e^{t} - 1} = \sum_{n=0}^{\infty} B_n(x) \frac{t^n}{n!},\label{eq19}
\end{equation}
with the first few polynomials given by:
\begin{align*}
B_0(x) &= 1, \\
B_1(x) &= x - \frac{1}{2}, \\
B_2(x) &= x^2 - x + \frac{1}{6}, \\
B_3(x) &= x^3 - \frac{3}{2}x^2 + \frac{1}{2}x. \label{eq20}
\end{align*}
Furthermore, for any positive integer $n$, the periodic Bernoulli function is defined as $\bar{B}_n = B_n(\{x\})$, where $\{x\}$ is the fractional part of $x$. The function $\bar{B}_n$ is periodic with period 1 and is continuous on it, implying that the fractional parts of the Bernoulli numbers are dense in $[0.1]$ \cite{Andrews1999}. This property, together with Elliott's result \cite{Elliot1998}, leads to the ultimate form of the partition function sum \cite{SiouaneLow2024,SiouaneTheo2024}:
\begin{equation}
\sum_{n=0}^{\infty} f(n) = \frac{f(0)}{2} + \int_{0}^{\infty} f(n)\, dn + \sum_{i=2}^{p} \frac{B_i\, f^{(i-1)}(0)}{i!} - \int_{0}^{1} \frac{B_p(1-t)}{p!} f^{(p)}(t)\, dt.
\label{eq:szp}
\end{equation}
The advantage of using the Euler-Maclaurin formula in this context is substantial \cite{SiouaneLow2024,SiouaneTheo2024}
By transforming the original sum into integral plus correction terms, we gain both analytical and computational efficiency. The integral often admits a closed form or can be evaluated numerically with high precision, while the correction terms systematically improve the approximation. This approach is compelling for partition functions, where direct summation may converge slowly or be impractical for large systems. The Euler-Maclaurin method thus provides a robust framework for extracting asymptotic behavior, estimating errors, and obtaining accurate results with minimal computational effort.
In the realm of statistical physics, a wide array of thermodynamic properties-including the partition function $Z(\beta)$, free energy $F(\beta)$, internal energy $U(\beta)$, entropy $S(\beta)$, and specific heat $C_v(\beta)$-can be systematically derived from the fundamental Boltzmann factor, $e^{-\beta E}$.   These quantities provide deep insight into the macroscopic behavior of a physical system based on its microscopic energy levels.   The relationships between these thermodynamic quantities and the partition function are given by : 
\begin{gather}
F(\beta) = -\frac{1}{\beta} \ln Z(\beta), \quad 
U(\beta) = -\frac{\partial \ln Z(\beta)}{\partial \beta}, \\
\frac{S(\beta)}{k_B} = \ln Z(\beta) - \beta \frac{\partial \ln Z(\beta)}{\partial \beta}, \quad 
\frac{C_v(\beta)}{k_B} = \beta^2 \frac{\partial^2 \ln Z(\beta)}{\partial \beta^2}.
\label{eq20}
\end{gather}

Here, $\beta = 1/(k_B T)$, where $k_B$ is Boltzmann’s constant and $T$ is the absolute temperature.   The partition function $Z(\beta)$ serves as the cornerstone of statistical mechanics, encapsulating the statistical weights of all accessible energy states and thereby governing the thermodynamic behavior of the system. 

It is essential to clarify the notation for the specific heat $C_v$.   The subscript $v$ in $C_v = \left(\frac{\partial U}{\partial T}\right)_{v=\text { constant }}$ indicates that the derivative is taken at constant $v$, which can refer to length (in 1D systems), area (in 2D systems), or volume (in 3D systems), depending on the dimensionality of the system under study.   This distinction is crucial, as the physical interpretation of $C_v$ adapts to the geometry and constraints of the system. 

These thermodynamic quantities not only characterize the equilibrium properties of the system, but also provide a foundation for understanding phase transitions, response functions, and fluctuations.   By expressing them in terms of derivatives of the partition function, we gain a unified and powerful framework for exploring the rich phenomenology of statistical systems across different dimensions and physical settings.\\

Before discussing the numerical findings, it is essential to outline the following considerations. To achieve an accurate evaluation of thermodynamic quantities, this study confines its analysis to stationary states characterized by positive energy, as referenced in \cite{Boumali2015}. This approach is substantiated through the following observations:
\begin{itemize}
    \item The Klein-Gordon equation, when incorporating interaction terms, facilitates a precise application of the Foldy-Wouthuysen transformation (FWT) \cite{Foldy1950,Foldy1952}. The Hamiltonian formulation of this equation was addressed by using the Feshbach-Villars transformation \cite{feshbach1958}. Khan and Jagannathan employed this methodology to investigate the quantum mechanics of charged particles in beam optics \cite{KhanJagannathan1994OperatorApproach}. Furthermore, Silenko \cite{Silenko2008} effectively applied the generalized Case-Foldy-Feshbach-Villars (CFFV) and Foldy-Wouthuysen transformations to derive the Hamiltonian for relativistic scalar particles within an electromagnetic field.
    \item This methodology ensures a clear distinction between positive and negative energy solutions, thereby preventing their intermixing. Notably, Myers \emph{et al.} \cite{Myers2021}, in their study of quantum Otto engines at relativistic energies, adopt the same FW-based rationale when deriving the thermal properties of relativistic oscillators such as the Dirac and Klein–Gordon oscillators. Although this reasoning is appropriate for the Dirac oscillator, they do not discuss its applicability to the Klein–Gordon oscillator, whose standard second-order formulation does not admit a first-order Hamiltonian representation without recasting via the Feshbach–Villars (and subsequent FW) transformation.  
    \item The evaluation of thermodynamic quantities was carried out exclusively using the positive energy spectrum, where $E_n > 0$. Incorporating negative energy states ($E_n < 0$) into the naive Gibbs sum, $Z = \sum_n e^{-\beta E_n}$, results in divergence. Consequently, for the single-particle canonical ensemble considered herein, the positive energy modes are sufficient to define a finite and physically meaningful partition function.
\end{itemize}

Moreover, to our knowledge, the first consideration of negative energy contributions in thermal calculations appears in the work of Myers et al.\cite{Myers2021}, who argued that negative energy modes act as supersymmetric counterparts to positive energy states.  Although this represents a novel conceptual advance, its direct application to relativistic oscillators produces an unphysical doubling of the specific heat.  Furthermore, in the nonrelativistic limit it predicts heat-capacity values twice those empirically observed in solid-state systems—where models that consider only positive-energy (electron) states already achieve excellent agreement with experimental measurements \cite{Boumalizna22015}. 

Based on these considerations, we focus our study on the thermal properties of the one-dimensional Klein-Gordon oscillator.  Particular emphasis is placed on the variation of the partition function and specific heat over this dimension, underscoring the impact of the positive-energy sector on the system's thermodynamic behaviour.

Now, to achieve this, the partition function $Z$ is evaluated using the following numerical methodology:
To evaluate the partition function $Z$ given in Eq. (\ref{eq:zp}), the Euler–Maclaurin summation formula (Eq. \ref{eq:fz}) is applied. This approach systematically approximates the discrete sum over energy levels by a continuous integral supplemented with a finite number of correction terms. Due to the intricate structure of the DSR-modified energy spectra $E_n$, the integral does not admit a closed analytical solution, although it converges rapidly for the relevant parameter range. Consequently, the integral $\int_0^\infty f(x)\,dx$, where $f(x) = \exp(-\beta E_x)$, is evaluated numerically using the trapezoidal rule to ensure high accuracy. For improved precision, we adopt $p = 3$ in the Euler-Maclaurin expansion, thus including correction terms up to the second derivative of $f(x)$ at the endpoints, with the appropriate Bernoulli numbers. Specifically, the calculation proceeds as follows: (i) $f(x)$ is constructed using the positive-energy sector of the corresponding DSR-modified spectrum; (ii) the integral is computed numerically with the trapezoidal method; and (iii) correction terms involving derivatives at the integration limits are incorporated up to order $p = 3$. This combined methodology ensures an accurate and robust evaluation of the partition function, forming the basis for the computation of all thermodynamic properties discussed in this work.
\\
\begin{figure*}
    \centering
    \subfigure[Partition function] {\includegraphics[scale=0.70]{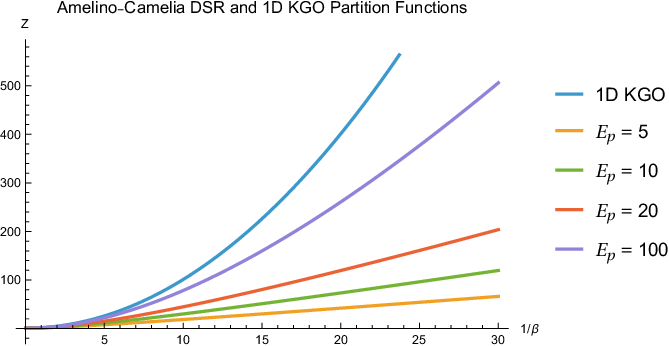}}\quad\subfigure[Specific heat]{\includegraphics[scale=0.70]{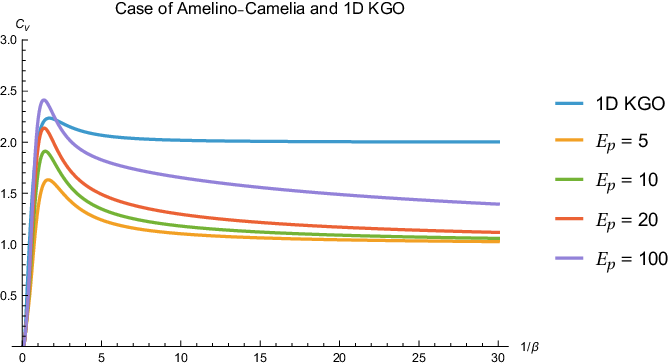}}\\
    \caption{ The behavior of the partition function and specific heat of the one-dimensional Klein-Gordon oscillator (1D KGO) as a function of $1/\beta$ in the framework of Amelino-Camelia DSR, and its comparison with the standard 1D KGO }
    \label{fig1}
\end{figure*}
\begin{figure*}
    \centering
    \subfigure[Partition function] {\includegraphics[scale=0.7]{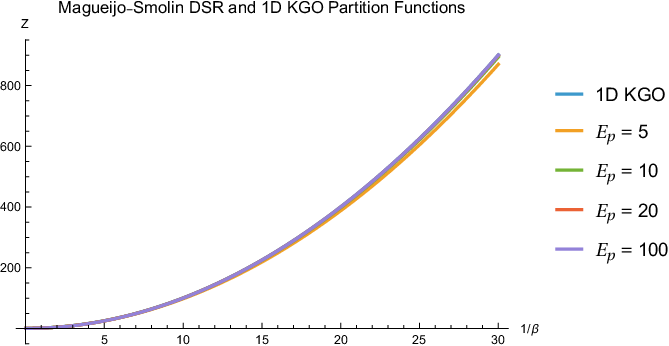}}\quad\subfigure[Specific heat]{\includegraphics[scale=0.70]{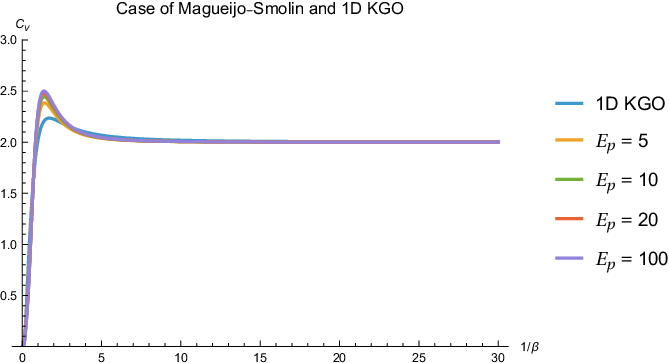}}\\
    \caption{The behavior of the partition function and specific heat of the one-dimensional Klein-Gordon oscillator (1D KGO) as a function of $1/\beta$ in the framework of Magueijo-Smolin Doubly DSR, and its comparison with the standard 1D KGO}
    \label{fig2}
\end{figure*}
\begin{figure*}
  \includegraphics[scale=1]{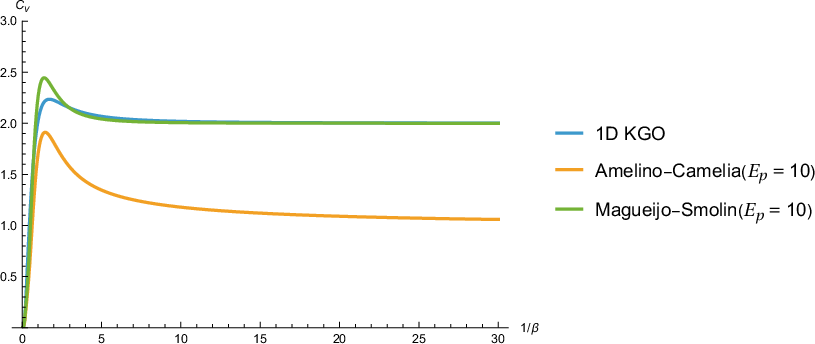}
  \caption{Comparison of the specific heat as a function of $1/\beta$ for the one-dimensional Klein-Gordon oscillator (1D KGO) in the frameworks of Amelino-Camelia and Magueijo-Smolin Doubly Special Relativity, with that of the standard 1D KGO.}
    \label{fig3}
\end{figure*}
The three figures explore the thermal properties of the one-dimensional Klein-Gordon oscillator (1D KGO) in the context of two different frameworks of DSR: the Amelino-Camelia DSR and Magueijo-Smolin DSR, compared with the standard 1D KGO.\\
In Figure. \ref{fig1}, the thermal properties of the 1D Klein-Gordon oscillator under the Amelino-Camelia DSR framework are studied by studying the partition function $Z$ and specific heat $C_v$ as functions of the inverse temperature ($1/\beta$). The partition function (panel a) grows with $1/\beta$, stabilizing at higher temperatures, and shows diverse behaviors based on the deformation parameter $E_p$, which encodes the Planck energy scale. This observation highlights how the DSR-modified dispersion relation alters the statistical properties of the system, particularly at higher temperatures where quantum gravitational effects become significant. The specific heat (panel b) grows with temperature, reaching a plateau whose value and position depend on $E_p$, with larger $E_p$ approaching the behavior of the classic relativistic oscillator. In contrast, lower $E_p$ values exhibit more pronounced Planck-scale effects, changing the specific heat curve. These results reveal that the Amelino-Camelia DSR framework brings crucial thermal response alterations, notably inside the positive-energy sector, which is physically relevant due to the separation of particle and antiparticle states guaranteed by the Foldy-Wouthuysen transformation.\\
Figure. \ref{fig2} illustrates similar results for the 1D KGO under the Magueijo-Smolin DSR framework. The partition function and specific heat are again given as functions of $1/\beta$ for various values of $E_p$, demonstrating trends similar to those in Figure 1. However, the precise shapes and values of the curves diverge, indicating the specific adjustments each DSR model adds to the energy-momentum relation. The Magueijo-Smolin corrections provide different thermodynamic signatures, particularly at lower $E_p$ values where Planck-scale physics dominates. The temperature dependence of the specific heat and its peak features differ from those reported in the Amelino-Camelia case, demonstrating that the choice of DSR framework leaves a unique mark on the system’s thermal observables.\\
Figure. \ref{fig3} compares the specific heat \( C_v \) of the positive-energy sector of the 1D KGO for both DSR frameworks (with \( k = 10 \)), alongside the normal 1D KGO. This comparison demonstrates that while both DSR-modified models diverge from the usual scenario, they do so in distinct manners. The specific heat curves for Amelino-Camelia and Magueijo-Smolin DSRs differ in both magnitude and temperature dependence. These fluctuations directly reflect the discrepancies in the DSR-modified dispersion relations, providing proof that thermodynamic parameters like specific heat can serve as sensitive markers of Planck-scale physics. The analysis is focused on the positive-energy sector, as the KGO ensures a precise separation of positive and negative energy solutions, guaranteeing that only physically relevant states contribute to the observed thermodynamic behavior.\\
\begin{figure*}
    \centering
    \subfigure[Entropy for AC model] {\includegraphics[scale=0.7]{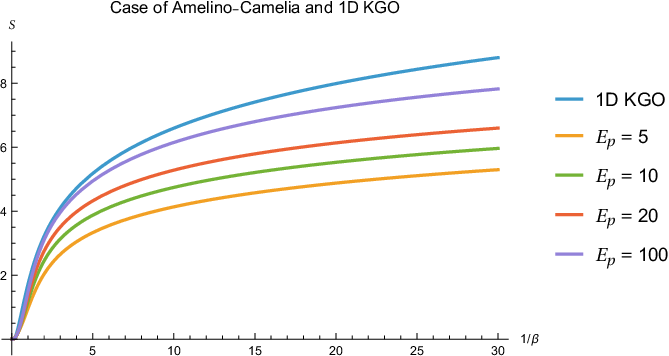}}\quad\subfigure[Entropy for MS model]{\includegraphics[scale=0.70]{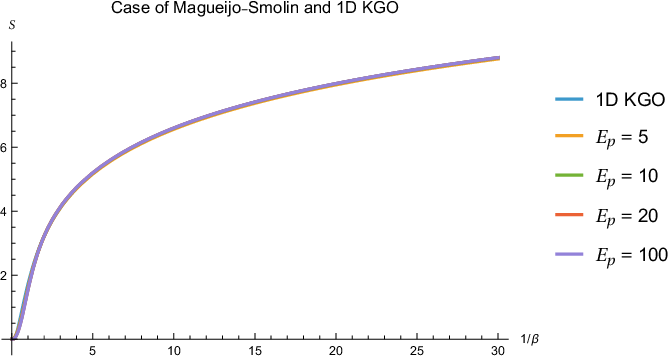}}\\
    \caption{The behavior of the entropy function of the one-dimensional Klein-Gordon oscillator (1D KGO) as a function of $1/\beta$ in the framework of both Amelino-Camelia and Magueijo-Smolin Doubly DSR, and its comparison with the standard 1D KGO}
    \label{fig4}
\end{figure*}
After those discussions, we are ready to inspect if the peaks that appear in the curves of specific heat are a sign of transition or not. For that, we have constructed the curve of the entropy functions.\\
We next analyse the numerical behaviour of the entropy $S(\beta)$ and specific heat $C_V(\beta)$ obtained from Eq. (41) with $\beta = 1/T$; the corresponding curves are shown in Fig. \ref{fig4}. 
The behaviour of the entropy $S(\beta)$ provides an immediate diagnostic of how the Amelino–Camelia (AC) and Magueijo–Smolin (MS) deformations filter down to macroscopic observables. For every value of the deformation scale $E_{\mathrm{p}}$ we considered, the entropy rises smoothly and monotonically with $1/\beta$ (i.e., with temperature) and exhibits no inflection points or plateaus. In the AC model, the additional linear-in-$n/E_{\mathrm{p}}$ term in the energy spectrum effectively compresses the level spacing at high quantum numbers, giving a slightly larger density of states and hence a slightly steeper $S(1/\beta)$ curve compared with the undeformed oscillator. The MS model, by contrast, alters the spectrum through a nonlinear rescaling of the mass term; this redistributes states more gently, leading to an entropy curve that tracks the standard Klein–Gordon oscillator almost in parallel once $E_{\mathrm{p}}\gtrsim10$. Crucially, neither model produces a peak, kink, or other non-analytic structure in $S(1/\beta)$.\\
Because the specific heat can be expressed as
\begin{equation} \label{eqcv}
C_{v}=T\,\frac{\partial S}{\partial T}
     =-\beta\,\frac{\partial S}{\partial\beta},  
\end{equation}

It is governed directly by the temperature derivative of the entropy.  When $\beta$ is analytic and strictly monotonic, $C_{v}$ remains positive definite and does not exhibit singular behavior, thus excluding any latent or continuous thermodynamic phase transition.  Consequently, the broad maxima observed in the numerical curves of $Cv(1/\beta)$ are most appropriately identified as Schottky anomalies—smooth, bell-shaped peaks that emerge in finite-level quantum systems when the thermal energy becomes comparable to the lowest excitation gap $\Delta E$. Specifically, once $k_BT \sim \Delta E$, the population rapidly redistributes from the ground state to the first excited state, producing a pronounced maximum in the specific heat $C_v(T)$. This feature does not signify a thermodynamic phase transition; rather, it reflects the discrete-level occupation inherent to finite systems, as exemplified by paramagnetic ions, two-level defects, and quantum dots.\\ 
Accordingly, while the Amelino–Camelia and Magueijo–Smolin deformations alter the rate at which the entropy grows with temperature, they leave its analytic structure unchanged.  Both DSR frameworks, therefore, preserve the qualitative, transition-free character of the thermodynamic response.\\
Finally, To the best of our knowledge, no existing experimental data yet allow a confrontation with our predictions; nevertheless, the present numerical results serve as a helpful guide to the expected behaviour of the system within both the Amelino-Camelia and Magueijo-Smolin DSR frameworks.
In conclusion, these figures collectively demonstrate that the altered dispersion relations significantly impact the thermal and statistical properties of the 1D Klein-Gordon oscillator in DSR frameworks. The partition function and specific heat are essential markers of how quantum gravitational corrections, represented by the deformation parameter $E_p$, appear in observable thermodynamic quantities. Comparing the Amelino-Camelia and Magueijo-Smolin models indicates that not only the presence but also the form of Planck-scale alterations can be determined by a careful investigation of thermal properties. This highlights the importance of relativistic quantum oscillators as theoretical tools for testing the phenomenological effects of quantum gravity-inspired modifications to the special theory of relativity.

\section{Conclusion}
This study examined the impact of Amelino–Camelia and Magueijo–Smolin DSR deformations on the thermal behavior of the one-dimensional Klein–Gordon oscillator, with analysis restricted to the exact positive-energy sector ensured by a Foldy–Wouthuysen transformation. In both frameworks, Planck-scale effects introduce discernible modifications to the partition function and specific heat, with their magnitude and functional form controlled by the deformation scale. An entropy-based diagnostic confirms that the thermodynamic curves remain smooth and monotonic; the broad maxima observed correspond to Schottky-type anomalies arising from the finite, discrete spectrum, rather than indicators of a phase transition. The absence of non-analytic features in the entropy rules out both latent and continuous transitions in the deformed oscillator. These results underscore the utility of thermodynamic observables as sensitive probes of Planck-scale physics, while indicating that the signatures appear as quantitative shifts rather than qualitative changes. Future work may extend the analysis to higher dimensions, alternative quantum-gravity settings, or experimental analogue platforms capable of resolving the predicted Schottky peaks and thereby constraining DSR parameters with laboratory precision.
\section{ Acknowledgment} 
We are grateful to the referee for their careful reading of the paper and for the detailed suggestions, which helped us improve considerably.\\
NJ has been funded by the Science Committee of the Ministry of Science and Higher Education of the Republic of Kazakhstan, Program No. BR24992759.

\bibliographystyle{unsrt}
\bibliography{referencearticle}

\end{document}